\documentclass[12pt, twoside]{article}
\usepackage{a4wide,amssymb,cite}
\usepackage{epsfig,axodraw}

\newcommand{\be}{\begin{equation}}
\newcommand{\ee}{\end{equation}}
\newcommand{\bea}{\begin{eqnarray}}
\newcommand{\eea}{\end{eqnarray}}

\begin{document}

\begin{titlepage}

\rightline{DESY 06-148}

\begin{centering}
\vspace{1cm}
{\large {\bf  Gauge Coupling Unification in a 6D $SO(10)$ Orbifold GUT}}\\

\vspace{1.5cm}

 {\bf Hyun Min Lee}
\\
\vspace{.2in}

 Deutsches Elektronen-Synchrotron DESY, 22603 Hamburg, Germany \\

\vspace{.1in}
(e-mail address: hyun.min.lee@desy.de)

\end{centering}
\vspace{2cm}

\begin{abstract}
\noindent
We consider the gauge coupling running in a six-dimensional
$SO(10)$ orbifold GUT model.
The bulk gauge symmetry is broken down to
the standard model gauge group with an extra $U(1)_X$
by orbifold boundary conditions
and the extra $U(1)_X$ is further broken through the $U(1)_{B-L}$ breaking
with bulk hyper multiplets.
We obtain the corrections of Kaluza-Klein massive modes to the running
of the gauge couplings and
discuss their implication to the successful gauge coupling unification.

\end{abstract}

\vspace{1cm}
%



\end{titlepage}

Grand Unified Theories(GUTs) have been revived recently in the models of
extra dimensions which are compactified on orbifolds,
the so called GUT orbifolds\cite{oguts,kkl}.
Thanks to orbifold boundary conditions in extra dimensions,
a GUT gauge symmetry can be broken down to the Standard Model(SM) gauge
group without the need of a GUT Higgs field in the large representation
and the doublet-triplet splitting problem can be solved easily.

On an orbifold $M/\Gamma$ with $M$ a compact manifold
and $\Gamma$ a point group,
there are fixed points which transform into themselves
under $\Gamma$. When the orbifolding breaks the gauge symmetry,
there are some of fixed points where the active gauge symmetry is reduced.
Although the non-universal gauge kinetic terms localized at the fixed points
can be introduced at tree level
and generated even by loop corrections\cite{nibb,gls,lee},
those effects may be ignored by making the strong coupling assumption
at the GUT scale with a large volume of extra dimensions\cite{naive}.
Thus, due to contributions coming from Kaluza-Klein(KK) massive modes,
the GUT orbifolds can provide a minimal setup to
predict the QCD coupling for a successful gauge coupling unification.

In this paper,
we consider the running of the gauge couplings
in the six-dimensional $SO(10)$ orbifold GUT model proposed in Ref.~\cite{abc}.
This is the minimal setup to break $SO(10)$ down to the SM gauge group
up to a $U(1)$ factor
only by orbifold boundary conditions without obtaining massless modes from
the extra component of gauge bosons.
We compute the threshold corrections due to KK massive modes
to the gauge coupling running for a number of hyper multiplets with arbitrary
parities.
By taking the 5D limit where the bulk gauge group becomes the Pati-Salam
$SU(4)\times SU(2)_L\times SU(2)_R$,
we show that the shape dependent term of the KK threshold corrections
gives rise to
the 5D power-like threshold corrections with non-universal coefficient.
In this paper,
focusing on the case that the logarithmic threshold
corrections are important, we discuss
about the possibility of having a large
volume of extra dimensions compatible
with the success of the gauge coupling unification in specific realizations
of the MSSM.

In our case, after the orbifolding, on top of the SM gauge group,
there is an extra $U(1)_X$ gauge symmetry which has to
be broken by a usual Higgs breaking of the $U(1)_{B-L}$\cite{abc42,abcflavor}.
In so doing, $\bf 16$ Higgs multiplets are introduced
in the bulk, so one ends up with extra color triplets as zero modes.
Although the extra color triplets can get masses of order the $B-L$
breaking scale $M_{B-L}$ at the fixed points,
they could give a large threshold correction to the gauge couplings.
We show that the KK threshold corrections can come with opposite sign to
the threshold corrections of the color triplets.
Thus, even if $M_{B-L}$ is much smaller than the GUT scale,
we can get the successful gauge coupling unification
due to the cancellation between the large threshold corrections.
In this case, the volume of extra dimensions can be large enough
for satisfying the strong coupling assumption.
We take some specific examples of embedding hyper multiplets
to show explicitly that this is the case for $M_{B-L}$ being smaller
than the compactification scale.
There are an extensive list of references\cite{so10}
where related discussions on the gauge coupling unification
have been done mainly in the context of a 5D $SO(10)$ orbifold GUT.

Two extra dimensions are compactified on a torus and they are
identified by a $Z_2$ reflection symmetry to make up a $T^2/Z_2$ orbifold.
For the extra coordinates $z=x^5+ix^6$, there are double periodicities
in extra dimensions such as $z\sim z+2\pi R_5\sim z+2i\pi R_6$.
Due to the orbifold action, there are four fixed points or branes,
$z_0=0$, $z_1=\pi R_5$, $z_2=i\pi R_6$ and $z_3=\pi R_5+i\pi R_6$.

A bulk vector multiplet is composed of a vector multiplet $V$ and an adjoint
 chiral multiplet $\Sigma$ in 4D ${\cal N}=1$ language.
In order to break the bulk gauge symmetry down to the SM gauge group,
we introduce a nontrivial boundary condition at each fixed point
for a bulk vector multiplet by the parity matrices\cite{abc},
\bea
P_i V(-z+z_i)P_i^{-1}&=&V(z+z_i), \\
P_i \Sigma(-z+z_i)P_i^{-1}&=&-\Sigma(z+z_i), \ \ i=0,1,2,3,
\eea
where
\bea
P_0&=&I_{10\times 10}, \\
P_1 &=& {\rm diag}(-1,-1,-1,1,1)\times \sigma^0, \\
P_2&=&  {\rm diag}(1,1,1,1,1)\times \sigma^2,
\eea
and $P_3=P_1P_2$ from the consistency condition on the orbifold.
Then, the parity operations $P_1, P_2$ break $SO(10)$ down to
its maximal subgroups, Pati-Salam group $SU(4)\times SU(2)_L\times SU(2)_R$
and Georgi-Glashow group $SU(5)\times U(1)_X$, respectively.
The parity operation $P_3$ also breaks $SO(10)$ down to flipped $SU(5)$ but it
is not an independent breaking.
Thus, the intersection of two maximal surviving subgroups
leads to $SU(3)_C\times SU(2)_L\times U(1)_Y\times U(1)_X$
as the remaining gauge group. This can be seen from the gauge bosons with
positive parities:
${\bf 45}$ is decomposed into $\bf (15,1,1)_++(6,2,2)_-+(1,3,1)_++(1,1,3)_+$
under $P_1$ (where $\pm$ indicate the parities)
and $\bf 24_{0,+}+10_{-4,-}+\overline {10}_{4,-}+1_{0,+}$ under $P_2$.
Then, finally, the extra $U(1)_X$ or $U(1)_{B-L}$
has to be broken further by the VEV of bulk or brane Higgs fields.

A bulk hyper multiplet is composed of two chiral multiplets with opposite
charges $(H,H')$ and it satisfies
the orbifold boundary conditions
\bea
\eta_i P_i H(-z+z_i)&=&H(z+z_i), \\
\eta_i P_i H'(-z+z_i)&=&-H'(z+z_i), \ \ i=0,1,2,3,
\eea
with $\eta^2_i=1$.
Here $\eta_0=1$ and $\eta_3=\eta_1\eta_2$, independent of the representation
of the hyper multiplet.
We consider
a set of hyper multiplets, $N_{10}$ $\bf 10$'s and $N_{16}$ $\bf 16$'s
satisfying $N_{10}=2+N_{16}$ for no irreducible
anomalies\cite{anomaly6d,anomaly6dlocal}.
We also note
that both $N_{10}$ and $N_{16}$ have to be even for the absence of localized
anomalies unless there are split multiplets
at the fixed points\cite{anomaly6dlocal}.
${\bf 10}=(H, G,H^c,G^c)$ is decomposed into
$\bf (6,1,1)_-+(1,2,2)_+$ under $P_1$
and $\bf 5_{-2,-}+{\bar 5}_{2,+}$ under $P_2$.
On the other hand, ${\bf 16}=(Q,L,U,E,D^c,N^c)$
is decomposed into
$\bf (4,2,1)_++({\bar 4},2,1)_-$ under $P_1$
and $\bf 10_{1,-}+{\bar 5}_{-3,+}+1_{5,+}$ under $P_2$.

In a 6D non-Abelian gauge theory on orbifolds,
where there is no orbifold breaking of the gauge symmetry,
the one-loop effective action for the gauge field has been obtained\cite{gls}.
The analysis has been extended to 6D GUTs with the orbifold breaking
of GUT symmetry\cite{lee}.
By using the general result in the latter analysis,
we study the running of the 4D effective gauge
couplings of the SM gauge group much below the compactification scale
in 6D $SO(10)$ GUTs.
After including all possible contributions,
the running of the low-energy gauge couplings
are governed in dimensional regularization by
\bea
\frac{4\pi}{g^2_{{\rm eff}, a}(k^2)}&=&
\frac{4\pi}{g^2_u}
+\frac{1}{4\pi}{\tilde b}_a\ln\frac{M^2_*}{M^2_{B-L}}
+\frac{1}{4\pi}b'_a\ln\frac{M^2_{B-L}}{k^2} \nonumber \\
&&-\frac{1}{4\pi}\bigg(\sum_{\pm\pm}b^{\pm\pm}_a L_{\pm\pm}
+\sum_{\pm\mp}b^{\pm\mp}_a L_{\pm\mp}\bigg)
+\frac{1}{2\pi}(\Delta^l_a+\Delta^{B-L}_a)\label{grun}
\eea
where $M_*$ is the 6D fundamental scale, $M_{B-L}$
is the $B-L$ breaking scale,
$g_u$ is the universal renormalized gauge
coupling\footnote{Although there are also
power-like threshold corrections in the cutoff regularization\cite{dienes,gls},
they don't contribute to the differential running of gauge couplings.
Nevertheless, the power-like contributions
may have the net effect of placing an upper limit on the possible volume of
the extra dimensions\cite{dienes2}.}
and $\Delta^l_a$ are corrections due to
renormalized gauge couplings localized
at the Pati-Salam and flipped $SU(5)$ fixed points.
$\Delta^{B-L}_a$ stands for the effect due to the modification of the
KK masses due to the $B-L$ breaking brane-localized mass terms.
Note further that $b'_a=(33/5,1,-3)$ is the beta function
in the MSSM as given below the $B-L$ breaking scale
while ${\tilde b}_a$ is the beta function above the $B-L$ breaking scale.
More importantly, $L_{\pm\pm}(L_{\pm\mp})$ are the logarithmic
KK threshold corrections with the corresponding beta functions
$b^{\pm\pm}_a(b^{\pm\mp}_a)$. These are a purely bulk contribution\cite{lee}.

Here we present the details of the beta functions in eq.~(\ref{grun}).
We split ${\tilde b}_a$ into ${\tilde b}_a=b_a-c_a+b^m_a$.
Here $b_a$ is the contribution from zero modes which are distributed
both in the bulk and at the fixed points\cite{gls,lee}.
It is given by
\be
b_a=b^V_a+b^{10}_a+b^{16}_a \label{bmassless}
\ee
with
\bea
b^V_a&=&(0,-6,-9), \\
b^{10}_a&=&\frac{1}{4}N_{10}(1,1,1)
+\frac{1}{4}\sum_{10}\eta^{10}_1(\frac{1}{5}, 1,-1), \label{b10}
\\
b^{16}_a&=&\frac{1}{4}(2N_{16}-\sum_{16}\eta^{16}_2)(1,1,1)
+\frac{1}{4}\sum_{16}\eta^{16}_1(-\frac{6}{5},2,0)
\nonumber \\
&&+\frac{1}{4} \sum_{16}\eta^{16}_1\eta^{16}_2(\frac{7}{5},-1,-1)
\label{b16}
\eea
where $\eta^{10}_i$ and $\eta^{16}_i$
with $(\eta^{10}_i)^2=(\eta^{16}_i)^2=1(i=1,2)$
are the parities for ${\bf 10}$ and $\bf 16$, respectively.
$c_a$ is the beta function for vector-like
massless modes which would get tree-level brane masses of order the GUT scale.
Moreover, $b^m_a$ is the beta function for the brane-localized fields.
Depending on the parities, we get the different logarithms for
the KK threshold corrections as
\bea
L_{++}&=& \ln\Big[4e^{-2}|\eta(iu)|^4uV M^2_*\Big],\\
L_{-+}&=&\ln\Big[\frac{e^{-2}}{4}\Big|\vartheta_1(\frac{1}{2}|iu)\Big|^4uV
M^2_*\Big],\\
L_{+-}&=&\ln\Big[\frac{e^{-2}}{4}\Big|\vartheta_1(-\frac{1}{2}iu|iu)\Big|^4uV
M^2_*\Big],\\
L_{--}&=&\ln\Big[\frac{e^{-2}}{4}
\Big|\vartheta_1(\frac{1}{2}-\frac{1}{2}iu|iu)\Big|^4uV M^2_*\Big]
\eea
where $u=R_6/R_5$, $V=4\pi^2 R_5R_6$, $\eta$ and $\vartheta_1$ are
the Dedekind eta function and the Jacobi theta function, respectively.
The beta function for KK massive modes is
\bea
&&b^{++}_a=\frac{1}{4}(-8+N_{10}+2N_{16})(1,1,1), \\
&&b^{-+}_a=\frac{1}{4}(\frac{12}{5},4,0)
+\frac{1}{4}\sum_{10}\eta^{10}_1(\frac{1}{5},1,-1)
+\frac{1}{4}\sum_{16}\eta^{16}_1(-\frac{6}{5},2,0), \label{b-+}\\
&&b^{+-}_a=\frac{1}{4}(2+\sum_{16}\eta^{16}_2)(-1,-1,-1), \\
&&b^{--}_a=\frac{1}{4}(\frac{38}{5},-2,-2)
+\frac{1}{4}\sum_{16}\eta^{16}_1\eta^{16}_2(\frac{7}{5},-1,-1).\label{b--}
\eea
Compared to eq.~(\ref{bmassless}), we obtain the relation between beta
functions as
\be
b_a=(0,-4,-6)+b^{++}_a+b^{-+}_a+b^{+-}_a+b^{--}_a
\ee
where the first term is due to the difference between the beta functions
of ${\cal N}=1$ vector multiplets and ${\cal N}=2$ vector multiplets
for the SM gauge group.
Consequently, from the beta functions (\ref{b10}), (\ref{b16}), (\ref{b-+})
and (\ref{b--}),
one can find that the part proportional to $\eta^R_1$ or $\eta^R_1\eta^R_2$
is non-universal. So, because of the orbifold actions
associated with Pati-Salam and flipped $SU(5)$ gauge groups,
both massless and massive mode contributions
can affect the differential running of the gauge couplings.

For a number of hyper multiplets with arbitrary parities,
we assume that both vector-like particles (getting
brane masses of order the GUT scale) and brane-localized particles
fill GUT multiplets, i.e. $c_a$ and $b^m_a$ are universal.
In this case, those particles do not affect the unification of the one-loop
gauge couplings.
Then, we get the general formula for the differential running
of gauge couplings as
\bea
\frac{1}{g^2_3}-\frac{12}{7}\frac{1}{g^2_2}+\frac{5}{7}\frac{1}{g^2_1}
&=&\frac{1}{8\pi^2}\bigg({\tilde b}\ln\frac{M_*}{M_{B-L}}
-\frac{1}{2}b^{-+}L_{-+}
-\frac{1}{2}b^{--}L_{--}+{\tilde\Delta}^l+{\tilde\Delta}^{B-L}\bigg)
\label{gutrel}
\eea
where
\bea
{\tilde b}
&=&\frac{9}{7}-\frac{9}{14}\sum_{10}\eta^{10}_1
-\frac{15}{14}\sum_{16}\eta^{16}_1+\frac{3}{7}\sum_{16}\eta^{16}_1\eta^{16}_2,
\\
b^{-+}
&=&-\frac{9}{7}-\frac{9}{14}\sum_{10}\eta^{10}_1
-\frac{15}{14}\sum_{16}\eta^{16}_1, \\
b^{--}
&=&\frac{12}{7}+\frac{3}{7}\sum_{16}\eta^{16}_1\eta^{16}_2.
\eea
Thus, we find a general relation between coefficients as
\be
{\tilde b}=\frac{6}{7}+b^{-+}+b^{--}.\label{relation}
\ee
Then, from eq.~(\ref{gutrel}) with the relation (\ref{relation}),
we find the deviation from the 4D SGUT prediction
of the QCD coupling at $M_Z$,
i.e.  $\Delta\alpha_s\equiv\alpha^{KK}_s-\alpha^{SGUT,0}_s$ as
\bea
\Delta \alpha_s(M_Z)
&\approx&-\frac{1}{2\pi}\alpha^2_s(M_Z)\bigg\{{\tilde b}
\ln\frac{M_*}{M_{B-L}}-({\tilde b}-\frac{6}{7})\ln(M_*\sqrt{V}) \nonumber \\
&&\quad-\frac{1}{2}b^{-+}\ln\Big[\frac{e^{-2}}{4}\Big|\vartheta_1(\frac{1}{2}|iu)
\Big|^4u\Big] \nonumber \\
&&\quad -\frac{1}{2}({\tilde b}-\frac{6}{7}-b^{-+})
\ln\Big[\frac{e^{-2}}{4}\Big|\vartheta_1(\frac{1}{2}-\frac{1}{2}iu|iu)
\Big|^4u\Big] \nonumber \\
&&\quad+{\tilde\Delta}^l+{\tilde\Delta}^{B-L}\bigg\}.\label{dev}
\eea
The first term corresponds to the contribution due to the extra particles
above the $B-L$ scale. The second term is the volume dependent
correction due to the KK massive modes while the third part containing
the theta functions is the shape dependent correction.
The last two terms ${\tilde \Delta}^l$ and ${\tilde\Delta}^{B-L}$ are
the effect of the brane-localized gauge couplings and the $B-L$ breaking
brane-localized mass terms, respectively.

Suppose to take the 5D limit with $u=R_6/R_5\gg 1$,
in which case the bulk gauge group becomes the Pati-Salam and
there remain only two fixed points
with the Pati-Salam group and the SM gauge group enlarged
with a $U(1)$ factor.
Then, since $|\vartheta_1(z|iu)|\sim 2e^{-\pi u/4}|\sin(\pi z)|$ for $u\gg 1$,
the shape dependent terms could give a significant effect on the gauge
coupling unification by the non-universal power-like threshold corrections
proportional to $u$ as in the case with the bulk VEV of extra components
of gauge bosons for a simple gauge group\cite{hebecker}.
In this case, the effective 5D gauge coupling($1/g^2_5=1/(g^2_4 R_6)$)
gets a power-like threshold
correction like $u/R_6\sim 1/R_5$ which is set
by the mass scale of heavy gauge bosons
belonging to $SO(10)/SU(4)\times SU(2)_L\times SU(2)_R$.

On the other hand, when $u\sim 1$,
the shape dependent term is subdominant compared
to the other logarithmic terms.
As can be shown explicitly in the specific models,
the last two terms can be also ignored by making a strong coupling
assumption and choosing the $B-L$ breaking scale to be smaller than the
compactification, respectively.
Then, the first two logarithms become a dominant contribution.
For ${\tilde b}(\tilde b-\frac{6}{7})>0$, we can see that
the individual logarithm can be large, being compatible with
the gauge coupling unification due to a cancellation.
We will focus on this possibility later on.
The case with the anisotropic compactification $u\gg 1$
will be discussed in detail elsewhere in Ref.~\cite{lee}.

\begin{figure}[t]

\begin{center}

\epsfig{file=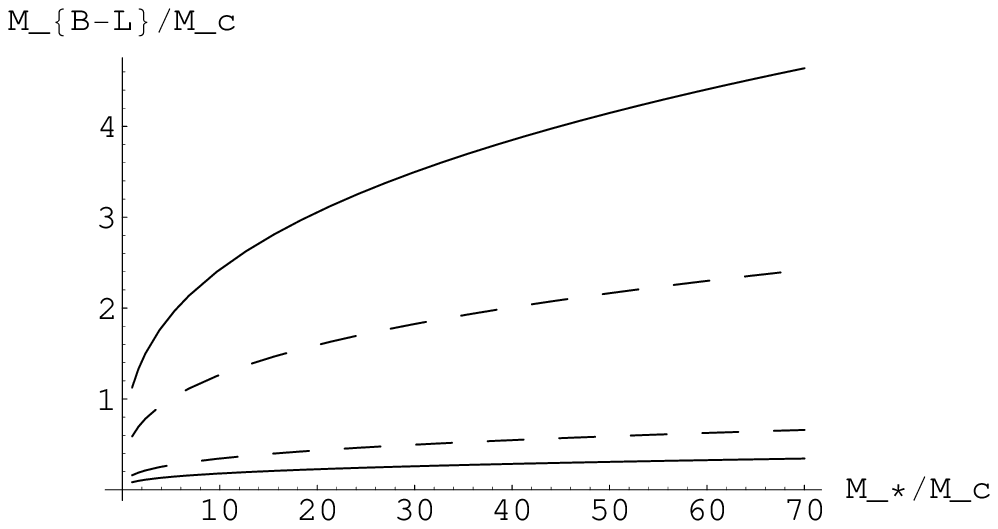,width=8cm,height=5cm}
\epsfig{file=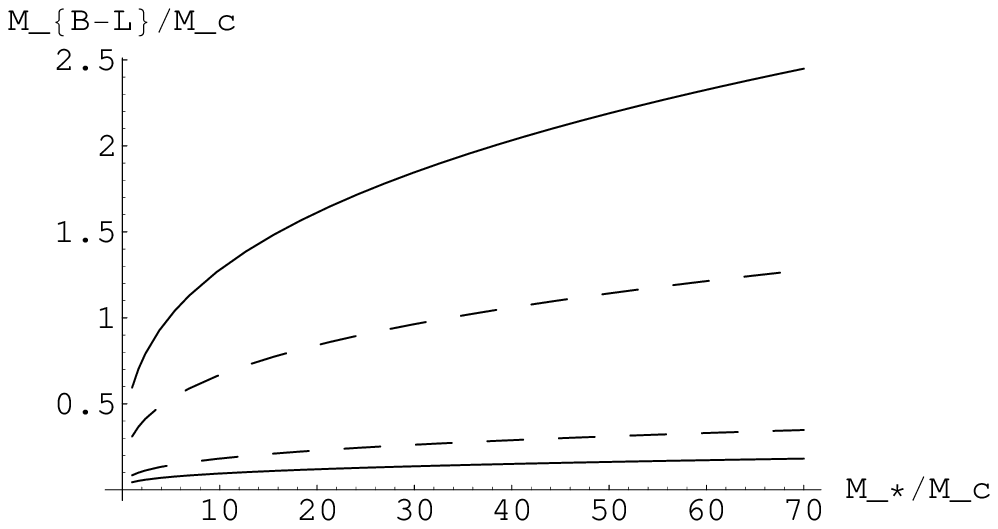,width=8cm,height=5cm}

\caption{The $1\sigma$ and $2\sigma$ band of $\Delta \alpha_s$: the model I
on the left and the model II on the right for $u=R_6/R_5\sim 1$.
The dashed lines and the thin lines denote $1\sigma$ and $2\sigma$
bounds of the experimental data, respectively.
}
\end{center}
\label{alpha}

\end{figure}

Now we are in a position
to apply our general formula (\ref{dev}) to particular cases
for the unification of the SM gauge couplings.
To this purpose, we consider some known $SO(10)$ models of embedding
the MSSM into the extra dimensions.
In the minimal model(: model I)\cite{abc42}
that contains Higgs fields in the bulk
for breaking $U(1)_{B-L}$ and the SM gauge group\footnote{In order to cancel the bulk anomalies due to one $\bf 45$, we need to add
in the bulk two ${\bf 10}$'s.
So, it is necessary to have two Higgs doublets of the $\bf 10$'s in the bulk
unlike in 5D case\cite{kkl}.
Moreover, in order to break the $U(1)_{B-L}$, we need one ${\bf 16}$
in the bulk. However, for cancellation of localized and bulk anomalies,
one needs
one $\bf \overline {16}$ and two more ${\bf 10}$'s.}, there are
4 ${\bf 10}$'s with parities $(\eta_1,\eta_2)$
such as $H_1=(+,+)$,
$H_2=(+,-)$, $H_3=(-,+)$ and $H_4=(-,-)$,
and one pair of $\bf 16$ and $\bf \overline {16}$
with parities $\Phi=(-,+)$, $\Phi^c=(-,+)$.
Then, the resulting massless modes are two doublet Higgs fields $H^c_1$ and $H_2$ from $H_1$ and $H_2$,
and $G^c_3,G_4, (D^c,N^c),(D,N)$ from $H_3,H_4,\Phi$ and $\Phi^c$ in order.
Moreover, each family of quarks and leptons is introduced as a $\bf 16$
being localized at the fixed point without $SO(10)$ gauge symmetry.
After the $B-L$ breaking via the bulk $\bf 16$'s
with $\langle N\rangle=\langle N^c\rangle\neq 0$,
neutrino masses are generated at the fixed points by a usual
see-saw mechanism. Moreover,
$G^c_3,G_4, (D^c,N^c),(D,N)$ can acquire masses
of order the $B-L$ breaking scale by the brane
superpotential\cite{abc42,abcflavor}
$W=\lambda N D G^c_3+\lambda' N^c D^c G_4$
for $\langle N\rangle=\langle N^c\rangle\neq 0$.
In this case, since $\sum_{10}\eta^{10}_1=0$, $\sum_{16}\eta^{16}_1
=\sum_{16}\eta^{16}_1\eta^{16}_2=-2$,
we get the values ${\tilde b}=\frac{18}{7},b^{-+}=b^{--}=\frac{6}{7}$
in eq.~(\ref{dev}).

We consider another 6D $SO(10)$ GUT model
where the realistic flavor structure of the SM
was discussed(: model II)\cite{abcflavor}.
In this case,
on top of the minimal model, there are more hyper multiplets:
2 ${\bf 10}$'s
such as $H_5=(-,+)$ and $H_6=(-,-)$,
and one pair of $\bf 16$ and $\bf \overline {16}$ with $\phi=(+,+)$
and $\phi^c=(+,+)$.
Then, there are additional zero modes
$G^c_5,G_6,L,L^c$ from $H_5,H_6,\phi$ and $\phi^c$ in order.
They are assumed to get brane masses of order the GUT scale.
Thus, the running
of gauge couplings between the GUT scale and the $B-L$ breaking scale
is the same as in the minimal model.
In this case,
since $\sum_{10}\eta^{10}_1=-2$, $\sum_{16}\eta^{16}_1
=\sum_{16}\eta^{16}_1\eta^{16}_2=0$,
we get the values ${\tilde b}=\frac{18}{7},b^{-+}=0$ and
$b^{--}=\frac{12}{7}$ in eq.~(\ref{dev}).

Consequently, in both cases, we can see that logarithmic contributions
of zero modes and those of KK massive modes appear with opposite signs
so that there is a possibility of having the large volume of extra
dimensions consistent with perturbativity and gauge coupling unification.
From the data of the electroweak gauge couplings at the scale of the $Z$ mass,
one can compare the predicted value of the QCD coupling in a theory to a
measure one\cite{expstrong} $\alpha^{exp}_s=0.1176\pm 0.0020$.
In the 4D supersymmetric GUTs, the prediction without
threshold corrections for the QCD coupling
is $\alpha^{SGUT,0}_s=0.130\pm 0.004$.
Thus, in this case, there is a discrepancy from the experimental data
as $\delta\alpha_s=\alpha^{SGUT,0}_s-\alpha^{exp}_s=0.0124\pm 0.0045$.
For the models that we considered above,
ignoring the unknown brane-localized
gauge couplings and the $B-L$ breaking effect,
we depict in Fig.~1 the parameter space of $(M_c,M_{B-L})$ with
$M_c\equiv 1/\sqrt{V}$ and $u\sim 1$, being compatible
with the experimental data.
Taking $M_*/M_c\sim 63/\sqrt{C}\sim 22$ with the group theory factor $C=8$ 
for strong coupling assumption at the 6D fundamental scale,
the correction due to the brane-localized gauge couplings
is ${\tilde \Delta}^l={\cal O}(1)$ so it is negligible to the KK threshold
corrections which is of order $\ln(M_*/M_c)\sim 3$.
For $M_{B-L}\ll M_c$, it has been shown\cite{dudas} that
the KK massive modes of the color triplets are modified to
$m^2_{n_5,n_6}\approx(n_5/2R_5)^2+(n_6/2R_6)^2+c M^2_{B-L}$ where
$c$ is of order unity independent of the KK level for $R_5\neq R_6$.
In this case,
the $B-L$ breaking effect to the differential running (\ref{gutrel}) is
estimated as ${\tilde\Delta}^{B-L}\sim M^2_{B-L}/M^2_c$. In the model I(II),
for $M_*/M_c\sim 22$,
$M_{B-L}/M_c$ can be as small as $0.23(0.12)$ at the $2\sigma$ level so that
the $B-L$ breaking can be suppressed compared to the KK threshold corrections.
Apart from the two models,
we can consider other possibilities of embedding the matter representations
into extra dimensions, like in the field-theory limit of a successful string
orbifold compactification\cite{buchmstring}
where there are two families at the fixed points and one family in the bulk.
In view of the general formula (\ref{dev}), however,
as far as an extra particle contributes to the running of the gauge couplings
above the $B-L$ breaking scale, $M_{B-L}$ tends to be close to $M_c$
for the success of the gauge coupling unification, independent of the details
of the model.

To conclude,
we have obtained the KK massive mode corrections
as a dominant contribution to the gauge coupling
running in a six-dimensional $SO(10)$ orbifold GUT model.
The shape dependent correction of the KK massive modes can be dominant
in the anisotropic compactification of the extra dimensions.
Compared to the 5D case, the 5D limit of our computation shows that
the 5D power-like threshold corrections can be computed
to be non-universal for the SM gauge couplings.
Focusing on the isotropic compactification of the extra dimensions,
we have shown that
there is a generic cancellation between the dominant logarithmic corrections
to the differential logarithmic running of the SM gauge couplings:
one is the contribution
of the extra particles above the $B-L$ scale
and the other is the KK massive mode contribution.
In the models that we considered, extra color triplets contribute to
the running of the gauge couplings above the $B-L$ scale
but the KK threshold corrections can be large enough
to cancel the contribution of the extra color triplets for the large volume
of extra dimensions. Therefore, the $B-L$ scale can be much smaller than
the GUT scale.

Since the $B-L$ breaking scale tends to be close to or larger than
the compactification scale as shown
in the allowed parameter space of Fig.~1, it may be also important to see
how much the modified KK massive modes of the color triplets
due to the brane-localized mass terms
can affect the running of the gauge couplings.
On the other hand, one can look for a consistent model
where the color triplets make up GUT multiplets together with some
extra doublets, i.e. ${\tilde b}=0$.
Then, the $B-L$ breaking would not be relevant for the gauge coupling
unification any more.
In this case, the extra dimensions could be also large enough
for the successful gauge coupling unification,
independent of the details of the model with hyper multiplets.
We leave the relevant issues in a future publication.

\section*{Acknowledgments}
The author would like to thank W.~Buchm\"uller,
L.~Covi and A.~Hebecker for useful comments and discussions.




\begin{thebibliography}{99}

\def\apj#1#2#3{Astrophys.\ J.\ {\bf #1}, #2 (#3)}
\def\ijmp#1#2#3{Int.\ J.\ Mod.\ Phys.\ {\bf #1}, #2 (#3)}
\def\mpl#1#2#3{Mod.\ Phys.\ Lett.\ {\bf A#1}, #2 (#3)}
\def\npb#1#2#3{Nucl.\ Phys.\ {\bf B#1}, #2 (#3)}
\def\plb#1#2#3{Phys.\ Lett.\ {\bf B#1}, #2 (#3)}
\def\prd#1#2#3{Phys.\ Rev.\ {\bf D#1}, #2 (#3)}
\def\prl#1#2#3{Phys.\ Rev.\ Lett.\ {\bf #1}, #2 (#3)}
\def\prt#1#2#3{Phys.\ Rep.\ {\bf #1}, #2 (#3)}
\def\sjnp#1#2#3{Sov.\ J.\ Nucl.\ Phys.\ {\bf #1}, #2 (#3)}
\def\zp#1#2#3{Z.\ Phys.\ {\bf #1}, #2 (#3)}
\def\jhep#1#2#3{JHEP\ {\bf #1}, #2 (#3)}
\def\ephjc#1#2#3{Europhys. J. C\ {\bf #1}, #2 (#3)}


\bibitem{oguts}
  Y.~Kawamura,
  Prog.\ Theor.\ Phys.\  {\bf 105} (2001) 999
  [arXiv:hep-ph/0012125];
  G.~Altarelli and F.~Feruglio,
  Phys.\ Lett.\ B {\bf 511} (2001) 257
  [arXiv:hep-ph/0102301];
  A.~B.~Kobakhidze,
  Phys.\ Lett.\ B {\bf 514} (2001) 131
  [arXiv:hep-ph/0102323];
  L.~J.~Hall and Y.~Nomura,
  Phys.\ Rev.\ D {\bf 64} (2001) 055003
  [arXiv:hep-ph/0103125];
  A.~Hebecker and J.~March-Russell,
  Nucl.\ Phys.\ B {\bf 613} (2001) 3
  [arXiv:hep-ph/0106166].

\bibitem{kkl}
  H.~D.~Kim, J.~E.~Kim and H.~M.~Lee,
  Eur.\ Phys.\ J.\ C {\bf 24} (2002) 159
  [arXiv:hep-ph/0112094].




\bibitem{nibb}
  S.~Groot Nibbelink and M.~Hillenbach,
  Phys.\ Lett.\ B {\bf 616} (2005) 125
  [arXiv:hep-th/0503153];
  Nucl.\ Phys.\ B {\bf 748} (2006) 60
  [arXiv:hep-th/0602155].


\bibitem{gls}
  D.~M.~Ghilencea, H.~M.~Lee and K.~Schmidt-Hoberg,
  JHEP {\bf 08} (2006) 009 [arXiv:hep-ph/0604215].


\bibitem{lee}
H.~M.~Lee, to be published.


\bibitem{naive}
  Z.~Chacko, M.~A.~Luty and E.~Ponton,
  JHEP {\bf 0007} (2000) 036
  [arXiv:hep-ph/9909248];
  L.~J.~Hall and Y.~Nomura,
  Phys.\ Rev.\ D {\bf 65} (2002) 125012
  [arXiv:hep-ph/0111068].


\bibitem{abc}
  T.~Asaka, W.~Buchm\"uller and L.~Covi,
  Phys.\ Lett.\ B {\bf 523} (2001) 199
  [arXiv:hep-ph/0108021].

\bibitem{abc42}
  T.~Asaka, W.~Buchm\"uller and L.~Covi,
  Phys.\ Lett.\ B {\bf 540} (2002) 295
  [arXiv:hep-ph/0204358].


\bibitem{abcflavor}
  T.~Asaka, W.~Buchm\"uller and L.~Covi,
  Phys.\ Lett.\ B {\bf 563} (2003) 209
  [arXiv:hep-ph/0304142].

\bibitem{so10}
  L.~J.~Hall, Y.~Nomura, T.~Okui and D.~R.~Smith,
  Phys.\ Rev.\ D {\bf 65} (2002) 035008
  [arXiv:hep-ph/0108071];
  R.~Dermisek and A.~Mafi,
  Phys.\ Rev.\ D {\bf 65} (2002) 055002
  [arXiv:hep-ph/0108139];
  H.~D.~Kim and S.~Raby,
  JHEP {\bf 0301} (2003) 056
  [arXiv:hep-ph/0212348];
  B.~Kyae, C.~A.~Lee and Q.~Shafi,
  Nucl.\ Phys.\ B {\bf 683} (2004) 105
  [arXiv:hep-ph/0309205];
  M.~L.~Alciati and Y.~Lin,
  JHEP {\bf 0509} (2005) 061
  [arXiv:hep-ph/0506130].


\bibitem{anomaly6d}
  J.~Erler,
  J.\ Math.\ Phys.\  {\bf 35} (1994) 1819
  [arXiv:hep-th/9304104];
  A.~Hebecker and J.~March-Russell,
  Nucl.\ Phys.\ B {\bf 625} (2002) 128
  [arXiv:hep-ph/0107039].

\bibitem{anomaly6dlocal}
  T.~Asaka, W.~Buchm\"uller and L.~Covi,
  Nucl.\ Phys.\ B {\bf 648} (2003) 231
  [arXiv:hep-ph/0209144].

\bibitem{dienes}
  K.~R.~Dienes, E.~Dudas and T.~Gherghetta,
  Phys.\ Lett.\ B {\bf 436} (1998) 55
  [arXiv:hep-ph/9803466];
  Nucl.\ Phys.\ B {\bf 537} (1999) 47
  [arXiv:hep-ph/9806292].

\bibitem{dienes2}
  K.~R.~Dienes, E.~Dudas and T.~Gherghetta,
  Phys.\ Rev.\ Lett.\  {\bf 91} (2003) 061601
  [arXiv:hep-th/0210294].


\bibitem{expstrong}
W.-M. Yao {\it et al.}[Particle Data Group Collaboration],
J. Phys. G {\bf 33} (2006) 1.

\bibitem{buchmstring}
  W.~Buchm\"uller, K.~Hamaguchi, O.~Lebedev and M.~Ratz,
  Phys.\ Rev.\ Lett.\  {\bf 96} (2006) 121602
  [arXiv:hep-ph/0511035];
  arXiv:hep-th/0606187.


\bibitem{dudas}
  E.~Dudas, C.~Grojean and S.~K.~Vempati,
  arXiv:hep-ph/0511001.

\bibitem{hebecker}
  A.~Hebecker and A.~Westphal,
  Annals Phys.\  {\bf 305} (2003) 119
  [arXiv:hep-ph/0212175];
  Nucl.\ Phys.\ B {\bf 701} (2004) 273
  [arXiv:hep-th/0407014].




\end{thebibliography}
\end{document}